\documentclass[twocolumn,showpacs,preprintnumbers,amsmath,amssymb]{revtex4}

\usepackage{graphicx}
\usepackage{dcolumn}
\usepackage{bm}

\begin{document}

\preprint{hep-th/0404242}

\title{Unstable D-branes and String
Cosmology\footnote{Proceedings of International Conference on Gravitation
and Astrophysics (ICGA6), 6-9 October 2003, Seoul, Korea.
Talk was given by Y. Kim.}
}

\author{Chanju Kim}
\email{cjkim@ewha.ac.kr}
\affiliation{Department of Physics, Ewha Womans University,
Seoul 120-750, Korea
}%

\author{Hang Bae Kim}
\email{hbkim@hanyang.ac.kr}
\affiliation{Department of Physics,
Hanyang University, Seoul 133-791, Korea
}%

\author{Yoonbai Kim, O-Kab Kwon, Chong Oh Lee}
\email{yoonbai@skku.edu, okab@skku.edu, cohlee@newton.skku.ac.kr}
\affiliation{BK21 Physics Research Division and Institute of Basic Science
Sungkyunkwan University, Suwon 440-746, Korea
}%

\begin{abstract}
Cosmological implication of rolling tachyons is reported in the context
of effective field theory. With a brief review of rolling tachyons in both
flat and curved spacetimes, we study the string cosmological model with
both tachyon and dilaton.
In the string frame, flat space solutions of both
initial-stage and late-time are obtained in closed form.
In the Einstein
frame, every expanding solution is decelerating.
When a Born-Infeld U(1) gauge field is coupled, enhancement of e-folding
of scale factor is also discussed by numerical analysis.
\end{abstract}

\maketitle

\section{Introduction}
Rolling tachyons have been proposed for the description of the homogeneous
decay of unstable D-brane in terms of both
boundary conformal field theory~\cite{Sen:2002nu} and
effective field theory~\cite{Sen:2002an}.
The role of rolling tachyons has been tested in
cosmology~\cite{Gibbons:2002md}, i.e., various topics include inflation,
dark matter, cosmological perturbation, and
reheating~\cite{Fairbairn:2002yp,Felder:2002sv,Mehen:2002xr,Kim:2002zr,
Gibbons:2003gb,Kim:2003qz,Mazumdar:2001mm}.
Though the tachyon driven cosmology is also a subject of string theory,
the basic language is effective field theory of the tachyon and graviton.
In this note, we will study the effects of linear dilaton from the bulk
and Born-Infeld type electromagnetic fields living on the brane
in addition to the two compulsory fields since the dilaton and Born-Infeld type
gauge fields are natural in string cosmology.
Another intriguing description of the rolling tachyons
in (1+1)-dimensional string theory is
the $c=1$ matrix model~\cite{McGreevy:2003kb}, however the cosmology
based on this language seems not realistic yet~\cite{Karczmarek:2003pv}.

The rest of this note is organized as follows. In section 2 and 3, we
review the rolling tachyons in both flat and curved spacetime.
In section 4, we consider the cosmology driven by rolling tachyons
in the context of string cosmology with dilaton. In section 5,
we find all possible rolling tachyon solutions in the presence of
Born-Infeld type electromagnetic fields. In section 6, cosmological
implication of such electromagnetic fields is studied in relation with
inflation. Section 7 is devoted to concluding remarks.

\section{Rolling Tachyons in Effective Theory}
Let us begin this section with
recapitulating the rolling tachyon
solution in the context of
effective field theory~\cite{Sen:2002an}.
Instability of an unstable D$p$-brane is described by the effective action
of tachyon $T$;
\begin{eqnarray}\label{act}
S&=&-{\cal T}_p \int d^{p+1}x\, V(T)
\sqrt{-\det(\eta_{\mu\nu}+\partial_{\mu}T\partial_{\nu}T)}
\nonumber \\
&=& -{\cal T}_p \int d^{p+1}x\, V(T)
\sqrt{1+\partial_{\mu}T\partial^{\mu}T} \, ,
\end{eqnarray}
where ${\cal T}_{p}$ is tension of the D$p$-brane.

Since tachyon potential measures variable tension of the
unstable D-brane, it should be a runaway potential connecting
$V(T=0)=1$ and $V(T=\infty)=0$. For the case of superstring,
${\rm Z}_{2}$-symmetry around its maximum at the origin is
assumed, $V(T)=V(-T)$.
Various forms of it have been proposed, e.g., $V(T)\sim e^{-T^{2}}$
with different derivative terms from
boundary string field theory~\cite{Gerasimov:2000zp} or $V(T)\sim e^{-|T|}$ for
large $|T|$ in Ref.~\cite{Sen:2002an}. In this paper, we employ the
form~\cite{Buchel:2002tj,Kim:2003he,Kim:2003qz,Leblond:2003db}
\begin{equation}\label{V3}
V(T)=\frac{1}{\cosh \left(\frac{T}{R}\right)}
\end{equation}
which connects the small and the large $T$ behaviors smoothly.
Here, $R$ is $\sqrt{2}$ for the non-BPS D-brane in the superstring
and 2 for the bosonic string.

Most of the physics of tachyon condensation is irrelevant to the detailed
form of the potential once it satisfies the runaway property and the
boundary values. So do the cosmological issues.
On the other hand, there are also the noteworthy features
with the 1/cosh potential (\ref{V3}) in the effective field theory:
(i)
Exact solutions are obtained for rolling
tachyon~\cite{Kim:2003he,Kim:2003ma} and tachyon kink
solutions on unstable D$p$ with a coupling of abelian gauge field
for arbitrary $p$~\cite{Lambert:2003zr,Kim:2003in,Kim:2003ma,Kim:2003uc}.
(ii)
This form of the potential has been derived in open string theory
by taking into account
the fluctuations around $\frac{1}{2}$S-brane configuration
with the higher derivatives neglected, i.e., $\partial^2 T = \partial^3 T=
\cdots = 0$~\cite{Kutasov:2003er,Okuyama:2003wm}. In addition,
the obtained classical solutions of the effective field theory
(\ref{act}) can be directly translated to those of the linearized tachyon
equation in open string
theory described by BCFT, i.e.,
an effective tachyon field $T(x)$ at on-shell and
a BCFT tachyon profile $\tau(x)$  has a one-to-one
correspondence~\cite{Kutasov:2003er}
\begin{equation}\label{oto}
\frac{\tau(x)}{R}=\sinh \left(\frac{T(x)}{R}\right).
\end{equation}
(iii)
Among the tachyon soliton solutions in the effective theory,
various tachyon array solutions of codimension one have been
found, namely, those formed by pure tachyon kink-antikink~\cite{Sen:1998tt,
Sen:1998ex,Lambert:2003zr,Kim:2003in}, tachyon kink-antikink
coupled to the electromagnetic field~\cite{Kim:2003in,Sen:2003bc,Kim:2003ma},
and tachyon tube-antitube~\cite{Kim:2003uc,Kim:2004xk}.
An interesting property of all these solutions is that
the periodicity of the array is independent of any integration constant
of the equation of motion only under Eq.~(\ref{V3})~\cite{Kim:2004xk}.
This periodic property of the array configurations in the effective
field theory
is desirable
if we wish to identify the array solution as a pair of
D$(p-1)\bar{{\rm D}}(p-1)$ obtained from an unstable
D$p$-brane wrapped on a cycle in the context of string
theory~\cite{Sen:1998ex,Horava:1998jy,Sen:2003bc}.

Let us discuss the rolling tachyon solutions in what follows.
Suppose a homogeneous configuration $T(t)$.
Then the Lagrangian density is given by
\begin{equation}\label{lagd}
{\cal L}=-{\cal T}_{p}V(T) \sqrt{1-\dot{T}^2},
\end{equation}
where the overdot $\dot{}$ denotes derivative of time.
The conjugate momentum density is
\begin{equation}
\Pi\equiv \frac{\partial {\cal L}}{\partial \dot{T}}
=\frac{{\cal T}_{p}V(T)\dot{T}}{
\sqrt{1-\dot{T}^2}},
\end{equation}
and Hamiltonian density is given by
\begin{equation}
{\cal H} = \sqrt{\Pi^2 + ({\cal T}_{p}V)^2}.
\end{equation}
Conservation of the Hamiltonian density $d{\cal H}/dt=\{ {\cal H}
,{\cal H} \}_{\mathrm{P.B.}}=0$
leads to constant energy density
\begin{equation}\label{econ}
{\cal H}=\rho=\frac{{\cal T}_{p}V}{\sqrt{1-\dot{T}^2}}
\end{equation}
which results in an integral equation
\begin{equation}\label{ineq}
\int \frac{dT}{\sqrt{1-({\cal T}_{p}V)^2/{\rho}^2}}
=t +\mathrm{constant}.
\end{equation}
For the specific form of tachyon potential (\ref{V3}), the integral
equation (\ref{ineq}) yields~\cite{Kim:2003he}
\begin{equation}\label{rtso}
\sinh\frac{T(t)}{R} = a_+ e^{t/R}-a_- e^{-t/R},
\end{equation}
where $a_\pm$ are related to the initial conditions as
\begin{eqnarray}
&a_\pm = \frac{1}{2}\left[\dot{T}_i \cosh (T_i/R) \pm \sinh (T_i/R)\right],&
\nonumber\\
& T_i \equiv T(0)~\mbox{and}~\dot{T}_i \equiv
\dot{T}(0).&
\end{eqnarray}
Due to time translation invariance and reflection symmetries ($T\leftrightarrow
-T$ and $t\leftrightarrow -t$), the rolling tachyon solution
(\ref{rtso}) is rewritten as a one-parameter family solution of ${\cal E}$
in a simpler form
\begin{equation}\label{rss2}
\left\{
\begin{array}{lccl}
\sinh\left(T_{{\rm RT}C}(t)/R\right)&=& \lambda_{{\rm RT}C}
\cosh\left(t/R\right) &
\mbox{for}~\rho<{\cal T}_{p} \\
\sinh(T_{\frac{1}{2}{\rm S}}(t)/R)
&=& \lambda_{\frac{1}{2}{\rm S}} \exp \left(t/R\right)
&\mbox{for}~\rho={\cal T}_{p}\\
\sinh\left(T_{{\rm RT}S}(t)/R\right)
&=& \lambda_{{\rm RT}S} \sinh\left(t/R\right)
&\mbox{for}~\rho>{\cal T}_{p}
\end{array}
\right. ,
\end{equation}
where
\begin{equation}\label{lam}
\left\{
\begin{array}{lccl}
\lambda_{{\rm RT}C}&=&\sqrt{({\cal T}_{p}/{\cal E})^{2}-1}
& \mbox{for}~\rho<{\cal T}_{p} \\
\lambda_{\frac{1}{2}{\rm S}}&=&1
& \mbox{for}~\rho={\cal T}_{p} \\
\lambda_{{\rm RT}S}&=&\sqrt{1-({\cal T}_{p}/{\cal E})^{2}}
&\mbox{for}~\rho>{\cal T}_{p}
\end{array}
\right. .
\end{equation}
Note that there are also trivial vacuum solutions, $T(t)=0$ for
$\rho={\cal T}_{p}$ and $T(t)=\pm\infty$
for $\rho=0$.
Therefore, through the relation (\ref{oto}),
the first rolling tachyon solution of
hyperbolic cosine form
$\tau_{{\rm RT}C}/R=\sinh\left(T_{{\rm RT}C}/R\right)$
coincides with the bounce solution in BCFT~\cite{Sen:2002nu},
the second exponential solution $\tau_{\frac{1}{2}{\rm S}}/R=
\sinh\left(T_{\frac{1}{2}{\rm S}}/R\right)$
for the critical value of energy density becomes
the $\frac{1}{2}$S-brane solution~\cite{Gutperle:2003xf},
and the third rolling tachyon solution of hyperbolic sine form
$\tau_{{\rm RT}S}/R=\sinh\left(T_{{\rm RT}S}/R\right)$
connects both vacua $T=\pm\infty$~\cite{Sen:2002nu}.

With the application to cosmology in the next section in mind, we note
the following facts which hold for any runaway tachyon potential:
(i) The energy density $\rho$
(\ref{econ}) is a constant of motion. At late time, it forces
\begin{equation}
T\rightarrow \infty,\qquad \dot{T}
\stackrel{t\rightarrow\infty}{\longrightarrow} 1.
\end{equation}
(ii) Pressure $P\equiv T^{i}_{\;i}/p$ is given by the Lagrangian density
(\ref{lagd}) so that it is negative and
approaches zero as time elapses, which
coincides with vanishing Lagrangian limit
\begin{equation}\label{press}
P={\cal L}=-{\cal T}_{p}V(T)\sqrt{1-\dot{T}^{2}}
\stackrel{t\rightarrow\infty}{\longrightarrow}0.
\end{equation}
{}From equation of state $P=w\rho$, we read
\begin{eqnarray}\label{rw1}
w=-(1-\dot{T}^{2})=-\frac{{\cal T}_{p}^{2}V(T(t))^{2}}{\rho^{2}}\le 0,
\end{eqnarray}
where $w=-{\cal T}_{p}^{2}/\rho^{2}$ at $T=0$ and $w\rightarrow 0$ at
$T\rightarrow \infty$.

\section{Cosmology of Rolling Tachyons}
An attractable topic of the rolling tachyon is its application
to cosmology as already indicated in~\cite{Sen:2002nu}.
The simplest setting for the cosmology of the
rolling tachyon is to assume a space-filling unstable D3-brane represented
by the field theoretic system composed of the graviton
$g_{\mu\nu}$ and the tachyon $T$ on the brane.
Among various research directions, we will review the results
of Ref.~\cite{Gibbons:2002md}.
The action is
\begin{equation}\label{gtac}
S=\int d^4x\sqrt{-g}\left(\frac{{\cal R}}{2\kappa^2}-{\cal T}_{3}V(T)
\sqrt{1+g^{\mu\nu}\partial_{\mu}T\partial_{\nu}T}\right).
\end{equation}

From the character of the tachyon potential in Eq.~(\ref{V3}), we immediately
read a few intriguing consequences of the rolling tachyon cosmology.
At initial stage
of the rolling tachyon, there exists a cosmological constant
$V(T\approx 0)\approx 1$ so that one expects easily solutions of inflationary
universe. At late time $T\rightarrow \infty$, we have tiny nonvanishing
but monotonically-decreasing cosmological constant at each instant, which
lets us consider it as a possible source of quintessence. Bad news for
comprising a realistic cosmological model may be the absence of reheating
and difficulty of density perturbation at late time due to disappearance
of perturbative degrees reflecting nonexistence of stationary vacuum point
in the monotonically decreasing tachyon potential.

For cosmological solutions, we try a spatially homogeneous and isotropic but
time-dependent configuration
\begin{equation}
ds^2=-dt^2+a^2(t)d\Omega^{2}_{k},\quad T=T(t),
\label{cosmos1}
\end{equation}
where $d\Omega^{2}_{k}$ corresponds, at least locally, to the metric of
a sphere ${\rm S}^{3}$, a flat Space ${\rm E}^{3}$, or a hyperbolic space
${\rm H}^{3}$ according to the value of $k=1,0,-1$, respectively.
Then the Einstein equations are summarized as
\begin{equation}\label{Eeq1}
\frac{\dot a^2}{a^2}+\frac{k}{a^2}
=\frac{\kappa^2}{3}\rho
= \frac{\kappa^2}{3}
\frac{{\cal T}_{3}V(T)}{\sqrt{1-\dot{T}^{2}}},
\end{equation}
\begin{equation}\label{Eeq2}
\frac{\ddot a}{a}
=-\frac{\kappa^2}{6}(\rho+3P)
=\frac{\kappa^2}{3}\frac{{\cal T}_{3}V(T)}{\sqrt{1-\dot{T}^{2}}}
\left(1-\frac{3}{2}\dot{T}^{2}\right),
\end{equation}
and the tachyon equation becomes
\begin{equation}\label{Teq}
\frac{\ddot T}{1-\dot T^2}+3\frac{\dot a}{a}\dot T
+\frac{1}{V}\frac{dV}{dT}=0.
\end{equation}

The tachyon equation (\ref{Teq}) is equivalent to conservation equation
of the energy-momentum tensor
\begin{equation}
\frac{\dot{\rho}}{\rho}=-3\frac{\dot{a}}{a}\dot{T}^{2}\le 0
\end{equation}
which  means that the energy density $\rho$ is no longer constant but
rather decreases in time. Since the pressure decreases rapidly to zero
\begin{equation}
P=-{\cal T}_{3}V(T){\sqrt{1-\dot{T}^{2}}}
\stackrel{t\rightarrow\infty}{\longrightarrow} 0,
\end{equation}
the equation of state $P=w\rho$ which is formally equivalent to the flat one
(\ref{rw1}) dictates $w\rightarrow 0$ for the gas of tachyon matter.

For the flat universe of $k=0$, cosmological evolution is rather simple
since both sides of Eq.~(\ref{Eeq1}) are positive semi-definite.
The second equation (\ref{Eeq2}) tells us that the expanding universe
is accelerating at onset of $T(t)\approx 0$ since the right-hand side of
Eq.~(\ref{Eeq2}) is also positive definite. As time $t$ goes, $\dot{T}$
exceeds $\sqrt{2/3}$ so that expansion of the universe slows down and then
finally the scale factor $a(t)$ will halt as
$a(t)\stackrel{t\rightarrow\infty}{\longrightarrow}{\rm constant}$.
This makes the second term of the tachyon equation (\ref{Teq}) vanish
at infinite time so that we confirm $\ddot{T}\rightarrow 0$ and
$\dot{T}\rightarrow 1$ as $T\rightarrow \infty$.

\section{String Cosmology of Rolling Tachyons}

In this section, we study the role of rolling tachyons in the cosmological
model with dilatonic gravity~\cite{Kim:2002zr}.
In the string frame, flat space solutions of
both initial-stage and late-time will be obtained in closed form.
In the Einstein frame, we will show that every expanding solution in flat space
is decelerating.

\subsection{String frame}\label{subsec:sfr}
We begin with a cosmological model induced from string theory, which
is confined on a D3-brane and
includes graviton $g_{\mu\nu}$, dilaton $\Phi$, and tachyon $T$.
In the string frame, the effective action of the bosonic sector of the D3-brane
system is given by
\begin{eqnarray}
S&=&\frac{1}{2\kappa^2}\int d^4x\sqrt{-g}\;e^{-2\Phi}\left({\cal R}
+4\nabla_\mu\Phi\nabla^\mu\Phi\right)
\nonumber\\
&&\hspace{-4mm}-{\cal T}_{3}\int d^4x\sqrt{-g}\;e^{-\Phi}\;V(T)
\sqrt{1+g^{\mu\nu}\partial_{\mu}T\partial_{\nu}T}\; ,
\label{act2}
\end{eqnarray}
where we turned off the Abelian gauge field $A_{\mu}$ on the D3-brane and
the second rank antisymmetric tensor fields $B_{\mu\nu}$ both on the brane and
in the bulk.

For cosmological solutions in the string frame, we assume in addition to
the metric ansatz (\ref{cosmos1})
\begin{equation}\label{Pt}
\Phi=\Phi(t).
\end{equation}
{}From the action (\ref{act2}), we read the equations in a simpler set
by introducing the shifted dilaton $\phi=2\Phi-3\ln a$
\begin{eqnarray}
\dot\phi^2-3H^2 +6\frac{k}{a^{2}}
&=& 2\kappa^{2}e^{\frac{\phi}{2}}a^{\frac{3}{2}}\rho_T, \label{metric4-1} \\
2(\dot{H}-H\dot\phi) +4\frac{k}{a^{2}}
&=& \kappa^{2} e^{\frac{\phi}{2}}a^{\frac{3}{2}} P_T, \label{metric4-2} \\
\dot\phi^2-2\ddot\phi +3H^2 +6\frac{k}{a^{2}}
&=& -\kappa^{2} e^{\frac{\phi}{2}}a^{\frac{3}{2}} P_T, \label{dilaton4} \\
\frac{\ddot T}{1-\dot T^2}+\frac12 (3H-\dot\phi)\dot T
&=&-\frac{1}{V}\frac{dV}{dT} ,
\label{tachyon4}
\end{eqnarray}
where $H=\dot a/a$ is the Hubble parameter, and
tachyon energy density $\rho_{T}$ and pressure $P_{T}$
defined by $T_{\;\;\;\;\nu}^{T\mu}\equiv
{\rm diag}(-\rho_{T},P_{T},P_{T},P_{T})$ are
\begin{equation}\label{pre}
\rho_T = {\cal T}_{3}\frac{V(T)}{\sqrt{1-\dot T^2}}~~~{\rm and}
~~~P_T = -{\cal T}_{3}V(T)\sqrt{1-\dot T^2},
\end{equation}
which formally coincide with Eq.~(\ref{econ}) and Eq.~(\ref{press}).
In the absence of detailed knowledge of $V(T)$,
we will examine characters of the solutions
based on the simplicity of tachyon equation of state
\begin{equation}
P_T=w_T\rho_T,\quad w_T=\dot T^2-1,
\label{T-eq-of-state}
\end{equation}
which is exactly the same as Eq.~(\ref{rw1}).
Note that $\sqrt{-g}$, or $a^3$ is not a scalar quantity
even in flat spatial geometry,
the shifted dilaton $\phi$ is not a scalar field in 3+1 dimensions.
The conservation equation is
\begin{equation} \label{conserv-eq2}
\dot\rho_T + \frac12 (3H-\dot\phi)\dot T^2 \rho_T=0.
\end{equation}
Now Eqs.~(\ref{metric4-1})--(\ref{dilaton4}) and Eq.~(\ref{T-eq-of-state})
are summarized by the following two equations
\begin{eqnarray}
2\ddot\phi-\dot\phi^2+2H\dot\phi-3H^2-2\dot H &=& -10\frac{k}{a^{2}},
\label{xx1} \\
w_T\dot\phi^2+4H\dot\phi-3w_TH^2-4\dot H &=& -(8+6w_{T})\frac{k}{a^{2}}.
\label{xx2}
\end{eqnarray}

Let us consider only the flat metric $(k=0)$ in the rest part of the paper.
If we express the dilaton $\phi$ as a function of the scale factor $a(t)$,
$\phi(t)=\phi(a(t))$, we can introduce a new variable $\psi$ such as
\begin{equation}
\psi\equiv a\phi' =\frac{\dot{\phi}}{H},
\end{equation}
where the prime denotes the differentiation with respect to $a$, and the
second equality shows that $\psi$ is the ratio between $\dot{\phi}$ and $H$.
Then Eqs.~(\ref{xx1}) and (\ref{xx2}) are combined into a single first-order
differential equation for $\psi$:
\begin{equation}\label{psi-eq}
4a\psi'+(\psi^2-3)(w_T\psi+2-w_T) = 0.
\end{equation}

{}From now on we look for the solutions of Eq.~(\ref{psi-eq}).
Above all one may easily find a constant solution
$\psi=\mp\sqrt{3}$ which is consistent with
Eqs.~(\ref{metric4-1})--(\ref{dilaton4}) only when $\rho_T=0$~:
\begin{eqnarray}
a(t)&=&a_{0}(1\mp\sqrt{3}H_{0}t)^{\mp 1/\sqrt{3}},\nonumber\\
H(t)&=&\frac{H_{0}}{1\mp\sqrt{3}H_{0}t},\label{ata}\\
\Phi(t)&=&\Phi_{0}-\frac{1\pm\sqrt{3}}{2}\ln(1\mp\sqrt{3}H_{0}t),
\label{pta}
\end{eqnarray}
where $H_{0}=H(t=0)$, $a_{0}=a(t=0)$, and $\Phi_{0}=\Phi(t=0)$
throughout this section. However, exactly-vanishing tachyon density
$\rho_{T}=0$ from Eq.~(\ref{metric4-1}) restricts
strictly the validity range of this
particular solution to that of vanishing tachyon potential, $V(T)=0$,
which leads to $T=\infty$.
The tachyon equation (\ref{tachyon4}) forces $\ddot{T}=0$ and $\dot{T}=1$ so
that the tachyon decouples $(w_{T}=P_{T}=0)$.
Therefore, the obtained solution (\ref{ata})--(\ref{pta}) corresponds to
that of string cosmology of the graviton and the dilaton before stabilization
but without the tachyon.

Since it is difficult to solve Eq.~(\ref{psi-eq}) with dynamical $w_{T}$,
let us assume that $w_T$ is time-independent
(or equivalently $a(t)$-independent).
We can think of the cases where
the constant $w_T$ can be a good approximation.
{}From the tachyon potential (\ref{V3}),
the first case is onset of tachyon rolling around the maximum point
and the second case is late-time rolling at large $T$ region.
In fact we can demonstrate that those two cases are the only possibility
as far as no singularity evolves.

When $w_T$ is a nonzero constant, Eq.~(\ref{psi-eq}) allows
a particular solution
\begin{equation}
\psi=\frac{w_T-2}{w_T}\equiv\beta.
\end{equation}
This provides a consistent solution of Eqs.~(\ref{metric4-1})--(\ref{dilaton4})
\begin{eqnarray}
a(t) &=& a_0\left(1+\frac{w_T^2+2}{2w_T}H_0t\right)^{\frac{2w_T}{w_T^2+2}},
\nonumber\\
H(t) &=&H_0\left(1+\frac{w_T^2+2}{2w_T}H_0t\right)^{-1},\\
\Phi(t)&=&\Phi_{0}+\frac{2(2w_{T}-1)}{w^{2}_{T}+2}\ln
\left(1+\frac{w^{2}_{T}+2}{2w_{T}}H_{0}t\right).
\label{ptw}
\end{eqnarray}
{}From Eq.~(\ref{metric4-1}) and Eq.~(\ref{conserv-eq2}),
the tachyon energy density $\rho_{T}$ is given by
\begin{eqnarray}\label{rhos}
\rho_T(t) &=& \frac{2-2w_T-w_T^2}{w_T^2\kappa^2e^{\Phi_0}}H_{0}^{2}
\nonumber\\
&&\times\left(1+\frac{w_T^2+2}{2w_T}H_0 t\right)^{-\frac{2(1+w_T)^2}{w_T^2+2}} .
\end{eqnarray}
Since the obtained solution is a constant solution of $\psi$,
it has only three initially-undetermined constants.
Specifically, the solution should satisfy $\dot\Phi=[(2w_T-1)/w_T]H$
so that the initial conditions also satisfy a relation
$\dot\Phi_0=[(2w_T-1)/w_T]H_0$.
Once we assume general solutions of $a(t)$-dependent $\psi$
with keeping constant nonzero $w_{T}$, they should be classified
by four independent parameters $(a_{0},H_{0},\Phi_{0},\dot{\Phi}_{0})$
instead of three in Eq.~(\ref{ptw}).

According to the aforementioned condition for valid $w_{T}$ values,
the obtained solution in Eq.~(\ref{ptw}) may be physically
relevant as the onset solution of $w_{T}=-1$ $(\psi=3)$.
In this case, $\rho_T(t)$ is reduced to a constant
$ \rho_T(t) = 3 e^{-\Phi_0} H_0^2 /\kappa^2 $.
Comparing this with the definition of $\rho_T$ in Eq.~(\ref{pre}),
the initial Hubble parameter $H_0$ is related to the dilaton as
$ H_0 = \pm \kappa e^{\Phi_0/2} \sqrt{T_3/3} $. Then, with $T(t) = 0$,
the tachyon equation of motion is automatically satisfied and hence
Eq.~(\ref{ptw}) becomes an exact solution of the
whole set of equations of motion (\ref{metric4-1})--(\ref{tachyon4}).
Since the tachyon field remains as constant at the maximum of the potential,
this solution describes the expanding or shrinking solution depending
on the initial Hubble parameter, with a constant vacuum energy corresponding
to brane tension due to tachyon sitting at the unstable
equilibrium point. Note that the interpretation as expanding or
shrinking solution needs to be more careful,
since we are working in the string frame.
Actually the behaviors are reversed in the Einstein frame as we will
see in the next subsection.

In order to study the behavior of the tachyon rolling down from the top of
the potential, now we slightly perturb this solution, i.e., look for a
solution with nonzero but small $T$ dependence.
So we treat $T$ as a small expansion parameter
and work up to the first-order in $T$.
Since the unperturbed solution satisfies $ 3H = \dot\Phi $,
the tachyon equation of motion (\ref{tachyon4}) becomes, to the
first-order in $T$,
\begin{equation}
\ddot{T} = -\frac{1}{V} \frac{dV}{dT} .
\end{equation}
This can easily be integrated to
\begin{equation}\label{teqn}
\frac12\dot{T}^2 = -\ln V + \mbox{constant} = \frac{T^2}{8\ln2} +
\mbox{constant},
\end{equation}
where we used the form of the potential near the origin (\ref{V3}).
Given the initial condition that $T=T_0$ and $\dot T = 0$ at $t=0$,
we can solve this equation (\ref{teqn}) and obtain
\begin{equation} \label{perturbed}
T(t) = T_0 \cosh{\alpha t},
\end{equation}
where $\alpha = 1/2\sqrt{\ln2}$.
Therefore tachyon starts to roll down the potential as a hyperbolic
cosine function.
Taking derivative, we find $\dot T = \alpha T_0 \sinh{\alpha t}$.
The range for which $\dot T$ remains small is then
$t\lesssim t_r\equiv 2\sqrt{\ln2} \sinh^{-1}(2\sqrt{\ln2}/T_0)$,
during which the approximation $w_T \simeq -1$ is good.
Unless the initial value $T_0$ is fine-tuned,
the tachyon follows the onset solution (\ref{ptw})--(\ref{rhos})
for $t\lesssim t_r$ and enters into rolling mode.

For more general solutions, the first-order differential
equation (\ref{psi-eq}) can be integrated to
\begin{equation}
a = C\left[\frac{\psi^2-3}{(\psi-\beta)^2}
\left(\frac{\psi-\sqrt3}{\psi+\sqrt3}\right)^{\frac{\beta}{\sqrt3}}
\right]^{\frac{\beta-1}{\beta^2-3}},
\end{equation}
where $C$ is an integration constant.
Note that this algebraic equation does not provide a closed form of $\psi$
in terms of the scale factor $a(t)$ except for a few cases, e.g.,
$w_T=0,-1/(\sqrt{3}-1/2),-2/(3\sqrt{3}-1)$.

Fortunately, for the late-time case of vanishing $w_T$,
we can obtain the solution in closed form
\begin{eqnarray}\label{cpm}
\psi = \sqrt3\ \frac
{C_+\left(\frac{a}{a_0}\right)^{\sqrt3/2}
+C_-\left(\frac{a}{a_0}\right)^{-\sqrt3/2}}
{C_+\left(\frac{a}{a_0}\right)^{\sqrt3/2}
-C_-\left(\frac{a}{a_0}\right)^{-\sqrt3/2}}.
\end{eqnarray}
Then the scale factor $a$ and the dilaton $\Phi$ are explicitly
expressed as functions of time $t$ by solving the equations
(\ref{metric4-2})--(\ref{dilaton4}):
\begin{eqnarray}
a(t) &=& a_0\left(\frac{C_-t+2}{C_+t+2}\right)^{1/\sqrt3},
\nonumber\\
H(t)&=&\frac{4H_0}{(C_-t+2)(C_+t+2)},
\label{att}\\
\Phi(t)&=&\Phi_0+\ln\left[
2\frac{(C_-t+2)^{(\sqrt3-1)/2}}{(C_+t+2)^{(\sqrt3+1)/2}}
\right],
\label{a-w0}
\end{eqnarray}
where $C_\pm=(3\mp\sqrt3)H_0-2\dot\Phi_0$.
We also read the tachyon density $\rho_T$ from Eq.~(\ref{metric4-1})
\begin{equation}\label{rhot}
\rho_T = C_+C_-e^{-\Phi_0}\
   \frac{(C_+t+2)^{(\sqrt3-1)/2}}{(C_-t+2)^{(\sqrt3+1)/2}}.
\end{equation}

Note that $C_{\pm}$ should have the same sign
from the positivity of the tachyon density (\ref{rhot}).
Let us first consider that both $C_{+}$ and $C_{-}$ are positive.
When $C_->C_+$ or equivalently $H_0>0$, the scale factor $a$ is growing
but saturates to a finite value
such as $a(\infty)=a_0 (C_-/C_+)^{1/\sqrt{3}}$ in the string frame.
When $C_-<C_+$, it decreases.
When $C_-=C_+$, $H_0=0$ so that the scale factor is a constant, $a(t)=a_0$.
For all of the cases, the dilaton $\Phi$ approaches negative infinity.
Note that $w_{T}=0$ means late-time, the tachyon density decreases like
$\rho_T\sim1/t$ as $t\rightarrow\infty$. Consistency check by using
Eq.~(\ref{conserv-eq2}) or equivalently by Eq.~(\ref{tachyon4}) provides us
the expected result, $\dot{T}\rightarrow1$.
If both $C_{+}$ and $C_{-}$ are negative, there appears a singularity at
finite time irrespective of relative magnitude of $C_{+}$ and $C_{-}$.

\subsection{Einstein frame}\label{subsec:Efr}
In the previous subsection, it was possible to obtain the cosmological solutions
analytically in a few simple but physically meaningful limiting cases.
To study the physical implications of what we found,
however, we need to work in the Einstein frame.
In this subsection, we will convert the cosmological solutions obtained
in the string frame to those in the Einstein frame
and discuss the physical behaviors.

First of all, let us rewrite the action (\ref{act2}) in the Einstein frame
of which metric
$g^E_{\mu\nu}$ is related to the string metric by a conformal transformation
\begin{equation}
g_{\mu\nu}=e^{2\Phi}g^E_{\mu\nu}.
\end{equation}
Note that we abbreviate the superscript $E$ for convenience in what follows.
Then the action (\ref{act2}) is changed to
\begin{eqnarray}
S&=&\frac{1}{2\kappa^2}\int d^4x\sqrt{-g}\left({\cal R}
-2\nabla_\mu\Phi\nabla^\mu\Phi\right)
\nonumber\\
&&\hspace{-7mm}-{\cal T}_{3}\int d^4x\sqrt{-g}\;e^{3\Phi}\;V(T)
\sqrt{1+e^{-2\Phi}\nabla_\mu T\nabla^\mu T}.
\label{act4}
\end{eqnarray}
Instead of deriving the equations from the action (\ref{act4}),
we obtain the same equations
from Eqs.~(\ref{metric4-1})--(\ref{tachyon4})
by using the metric of the form
\begin{equation} \label{einsteinmetric}
ds^2=e^{2\Phi}(-dt^2+a^2(t)d\Omega^{2}_{k}),
\end{equation}
and hence the time $t$ and the scale factor $a$ are related to
those in the string frame as
\begin{equation}\label{rel}
a_s=ae^{\Phi},\qquad dt_s=e^{\Phi}dt.
\end{equation}
Note that in this subsection
all the quantities are in the Einstein frame
except the variables with subscript $s$ which denote the quantities
in the string frame.
Then the Einstein equations for the flat
case $(k=0)$ in the string frame (\ref{metric4-1})--(\ref{metric4-2})
are converted to
\begin{eqnarray}
H^2 &=& \frac13 \dot\Phi^2 + \frac13 \kappa^2 e^{3\Phi}\rho_T\,,
           \label{einstein5}\\
\dot H &=& - \dot\Phi^2 - \frac12 \kappa^2 e^{\Phi}\rho_T \dot T^2\,,
           \label{einstein6}
\end{eqnarray}
where
tachyon energy density $\rho_T$ and pressure $P_T$ in the Einstein frame
are obtained by the replacement $\dot{T}_{s}=e^{-\Phi}\dot{T}$
in Eq.~(\ref{pre}),
\begin{eqnarray}
\rho_T &=& {\cal T}_{3}\frac{V(T)}{\sqrt{1-e^{-2\Phi}\dot T^2}}\nonumber\\
P_T &=& -{\cal T}_{3}V(T)\sqrt{1-e^{-2\Phi}\dot T^2} ,
\label{rpt}
\end{eqnarray}
and thereby $w_{T}$ is given as
\begin{equation}
w_{T}\equiv P_T /\rho_T=e^{-2\Phi}\dot T^2 -1.
\end{equation}

Demanding constant $w_{T}$ is nothing but asking a strong
proportionality condition between the dilaton and tachyon,
$\dot{T}\propto e^{\Phi}$.
Note that the pressure $P_{T}$ as shown in Eq.~(\ref{pre}) is always
negative irrespective
of both specific form of the tachyon potential $(V(T)\ge 0)$ and the value of
the kinetic term $(e^{-2\Phi}\dot{T}^{2}\le 1)$, and the value of $w_T$
interpolates smoothly between $-1$ and 0.

First we observe that the right-hand side of Eq.~(\ref{einstein5}) is always
positive, which means that the Hubble parameter $H(t)$ is either
positive definite or negative definite for all $t$ and it cannot change
the sign in the Einstein frame.
Let us first consider the case of positive Hubble parameter, $H(t)>0$.
Eq.~(\ref{einstein6}) shows $\dot H$ consists of two
terms both of which are negative definite for all $t$.  Since $H>0$ by
assumption, the only consistent behavior of $H$ in this case is that
$\dot H$ vanishes as $t\rightarrow \infty$, which, in turn implies that
$\dot\Phi$ and $e^{\Phi}\rho_T \dot T^2$ go to zero, separately.
It also means that $H$ should be a regular function for all $t$. In order to
find the large $t$ behavior of $H(t)$, one has to study $e^{-\Phi} \dot T$
in large $t$ limit which appears in the definition of $w_T$ in the
Einstein frame. Knowing that the functions are regular, it is not difficult
to show that the only possible behavior is $e^{-\Phi} \dot T \rightarrow 1$
as $t \rightarrow \infty$ after some straightforward analysis of
Eqs.~(\ref{einstein5})--(\ref{einstein6}). Combining it with the fact that
$\dot\Phi$ and $e^{\Phi}\rho_T \dot T^2$ vanish, we can immediately
conclude from Eq.~(\ref{einstein5}) that $H(t)$ should go to zero in large
$t$ limit.

The asymptotic behavior of fields in case of the positive Hubble parameter
can be found from the solution (\ref{a-w0}) since $w_T$ is essentially
zero for large $t$ as we just have seen above. The only thing to do is
to transform the expressions in the string frame to those in the
Einstein frame, using the relation (\ref{rel}).
Therefore, for large $t$, we find
\begin{eqnarray} \label{einstein_a}
a(t_s) &=& a_s(t_s)e^{-\Phi(t_s)} \nonumber \\
&\simeq&\frac{1}{2}a_{s0}e^{-\Phi_{0}}
(C_{+}t_s+2)^{\frac{\sqrt3+1}{2\sqrt3}}
(C_{-}t_s+2)^{\frac{\sqrt3-1}{2\sqrt3}}, \nonumber \\
t &=& \int dt_s e^{-\Phi}  \nonumber \\
   & \simeq & 2 e^{-\Phi_0} \int dt_s
       \frac{(C_+ t_s+2)^{(\sqrt3+1)/2}}{(C_- t_s+2)^{(\sqrt3-1)/2}}.
\end{eqnarray}
One can also identify the initial Hubble parameter $H_0$ in terms of $C_\pm$ as
\begin{equation}
H_0 = \frac14 e^{\Phi_0}\left[
               \left( 1 - \frac1{\sqrt3} \right) C_{-}
             + \left( 1 + \frac1{\sqrt3} \right) C_{+} \right].
\end{equation}
Note that $C_\pm$ have the same sign as the Hubble parameter $H$.
Now, with $C_\pm>0$, one can easily confirm from the above equation
(\ref{einstein_a}) that all the functions indeed behave regularly.
In $t_s\rightarrow\infty$ limit, $a\sim t_s$ and $t \sim t_s^{2}$
so that the asymptotic behavior of the scale factor becomes $a\sim t^{1/2}$.
This power law expansion in flat space is contrasted with the result of
Einstein gravity without the dilaton $\Phi$, where ultimately the scale
factor ceases to increase, $\lim_{t\rightarrow\infty}a(t)\rightarrow$ constant.
The behavior of tachyon density $\rho_T$ can be read from
Eq.~(\ref{rhot}) with $t$ replaced by $t_s$, which shows that
$\rho_T \sim t^{-1/2}$.
Since $w_T$ also goes to zero, the fluid of condensed tachyon becomes
pressureless.
Differently from ordinary scalar matter where matter domination
of pressureless gas is achieved
for the minimum kinetic energy $(\dot{T}\rightarrow 0)$, it is
attained for the maximum value of time dependence $(e^{-\Phi}\dot{T}
\rightarrow 1$ as $T\rightarrow \infty)$ for the tachyon potential
given in Eq.~(\ref{V3}).

When the Hubble parameter $H$ is negative, the situation is a bit more
complicated. Since $\dot H<0$ always,
$H$ becomes more and more negative and there is a possibility that
eventually $H$ diverges to negative infinity at some finite time.
Indeed, it turns out that all solutions in this case develop a singularity
at some finite time at which $H \rightarrow -\infty$ and $a \rightarrow 0$.
These big crunch solutions may not describe viable universes
in the sense of observed cosmological data.
Depending on initial conditions, the dilaton $\Phi$ diverges to either $\infty$ or
$-\infty$ and $\dot T$ goes to either $\infty$ or zero with the factor
$e^{-\Phi} \dot T$ remaining finite. It is rather tedious and not much
illuminating to show this explicitly, so here we will just content
ourselves to present a simple argument to understand the behavior.
Since the tachyon field $T$ rolls down from the maximum of the potential
to the minimum at infinite $T$, it is physically clear
that $ \dot T_s = e^{-\Phi} \dot T$ would eventually go to one unless
there is a singularity at some finite time. Suppose that there appeared
no singularity until some long time had passed so that $e^{-\Phi} \dot T$
approached to one sufficiently closely. Then Eq.~(\ref{einstein_a}) should be
a good approximate solution in this case. However, we know that both
$C_\pm$ are negative when $H<0$ and Eq.~(\ref{einstein_a}) is clearly
singular in this case. We have also verified the singular behavior for
various initial conditions using numerical analysis.

As mentioned in the previous section, the tachyon $T$ is decoupled when
$e^{-\Phi}\dot{T}=1$ and $T=\infty$. In this decoupling limit, characters of
the Einstein equations (\ref{einstein5})--(\ref{einstein6}) that $H^{2}>0$
and $\dot{H}<0$ do not change so that all the previous arguments can be applied.
Well-known cosmological solution of the dilaton gravity before stabilization
of the dilaton is
\begin{eqnarray}
a(t)&=&a_0 (1+3H_{0}t)^{1/3},~~~H(t)=\frac{H_{0}}{1+3H_{0}t},
\label{adg}\\
\Phi(t)&=&\Phi_{0}\pm\frac{1}{\sqrt{3}}\ln(1+3H_{0}t),
\label{pdg}
\end{eqnarray}
where the $(\pm)$ sign in Eq.~(\ref{pdg}) is due to the reflection symmetry
$(\Phi\leftrightarrow -\Phi)$ in the equations
(\ref{einstein5})--(\ref{einstein6}).
This solution can also be obtained throughout a transformation (\ref{rel})
from Eqs.~(\ref{ata})--(\ref{pta}).
For $H_{0}<0$, it is a big crunch solution $(a\rightarrow 0)$ which encounters
singularity $(H\rightarrow\infty,~\Phi\rightarrow \mp\infty)$ as $t\rightarrow
1/3|H_0|$. For $H_{0}>0$, it is an expanding but decelerating solution.
Since $a\sim t^{1/3}$, the power of expansion rate is increased from $1/3$ to
$1/2$ by the tachyonic effect as expected.

So far we discussed generic properties and asymptotic behaviors of solutions
in the Einstein frame. Now we consider the behavior at the onset.
The solution (\ref{ptw}) obtained by assuming constant $w_T$ is
transformed to the Einstein frame as
\begin{eqnarray}
a(t) &=& a_0\left[1+\frac{(w_T-2)^2}{2(1-w_T)}H_0t
        \right]^{\frac{2(1-w_T)}{(w_T-2)^2}}, \nonumber \\
e^{\Phi(t)} &=& e^{\Phi_0}\left[1+\frac{(w_T-2)^2}{2(1-w_T)}H_0t
        \right]^{\frac{2(2w_T-1)}{(w_T-2)^2}},
\end{eqnarray}
where the initial Hubble parameter $H_0$ is related to that in the string
frame by $H_0=e^{\Phi_0}H_{0s}(1-w_T)/w_T$.
Note that $H_0$ and $H_{0s}$ have opposite signs since $w_T<0$. Therefore
the expanding (shrinking) solution in the string frame corresponds to the
shrinking (expanding) solution in the Einstein frame.
For the onset solution with $w_T=-1$, the tachyon energy density $\rho_T$
is a constant as before, $\rho_T(t) = 3e^{-3\Phi_0} H_0^2/4 \kappa^2$.
Then the initial Hubble parameter is given by
$H_0 = \pm 2\kappa e^{3\Phi_0/2} \sqrt{{\cal T}_3/3}$, which describes the exact
solution that tachyon remains at the origin as explained in section 2.
Under a small perturbation tachyon starts
rolling down according to Eq.~(\ref{perturbed}) with $t$ replaced by
$t_s$. The rest of the discussion on the rolling behavior is the same as
in the string frame and the details will not be repeated here.

In conclusion the cosmological solution can be classified into two categories
depending on the value of the Hubble parameter $H(t)$ in the Einstein frame.
When the initial Hubble parameter $H_0$ is positive, the solution is
regular and the universe is expanding but decelerating as
$a(t) \sim \sqrt{t}$ while $e^{\Phi(t)}$ vanishes. When $H_0$ is negative,
there appears a singularity at some finite time $t$ at which the universe
shrinks to zero.

\section{Rolling Tachyons Coupled to U(1) Gauge Fields}
In this section we introduce the system of tachyon coupled to an Abelian
gauge field of Born-Infeld type and find the most general homogeneous
solution which
turns out to be constant electric and magnetic fields
together with rolling tachyon configuration~\cite{Kim:2003qz,Kim:2003ma}.

The unstable flat D$p$-brane system is described by
the following Born-Infeld type action
\begin{equation}\label{pfa}
S= -{\cal T}_{p} \int d^{p+1}x\; V(T) \sqrt{-\det (\eta_{\mu\nu} +
\partial_\mu T\partial_\nu T + F_{\mu\nu})}\, ,
\end{equation}
where $F_{\mu\nu}$ is field strength tensor of Abelian
gauge field $A_{\mu}$ on the D$p$-brane,
$F_{\mu\nu}=\partial_{\mu}A_{\nu}-\partial_{\nu}A_{\mu}$.

If we are interested in spatially homogeneous configurations of
\begin{equation}
T=T(t),\qquad F_{\mu\nu}=F_{\mu\nu}(t),
\end{equation}
then the proof in section 3 of Ref.~\cite{Kim:2003ma} tells us that
every component of
the field strength tensor $F_{\mu\nu}$ should be constant in order to satisfy
both Born-Infeld type equations of motion for the gauge fields
and Bianchi identity.
Then the action (\ref{pfa})
is rewritten as
\begin{equation}\label{siac}
S=-{\cal T}_{p}\int d^{p+1}x\; V(T)\sqrt{\beta_{p}-\alpha_{p0}\dot{T}^{2}},
\end{equation}
where $\alpha_{p0}$ is 00-component of the cofactor of matrix
$(X)_{\mu\nu}=\eta_{\mu\nu} +
\partial_\mu T\partial_\nu T + F_{\mu\nu}$
\begin{eqnarray}
\alpha_{p0}&=&C^{00}\ge 1,\\
\beta_{p}&=&-\det (\eta_{\mu\nu}+F_{\mu\nu}).
\end{eqnarray}
Since $\alpha_{p0}$ is positive, reality
condition of the action (\ref{siac}) requires positivity of $\beta_{p}$
either.
If we rescale the time variable $t$ as
\begin{equation}\label{tim}
\tilde{t}=\frac{t}{\sqrt{\alpha_{p0}/\beta_{p}}},
\end{equation}
the form of action (\ref{siac}) becomes the same as that of the pure tachyon
\begin{equation}\label{ppab}
S=-\tilde{{\cal T}}_{p}
\int d\tilde{t}\int d^{p}x\, V(T)\sqrt{1-\tilde{\dot{T}}^{2}},
\end{equation}
where $\tilde{{\cal T}}_{p}={\cal T}_{p}\sqrt{\alpha_{p0}}$ and
$\tilde{\dot{T}}=dT/d\tilde{t}$.
Subsequently, equation of motion is given by the conservation of energy
(Hamiltonian) density as given previously in Eq.~(\ref{econ})
\begin{equation}\label{eqn}
\tilde{\rho}\equiv \sqrt{\frac{\beta_{p}}{\alpha_{p0}}}
{\cal H}=\frac{\tilde{{\cal T}}_{p}V(T)}{\sqrt{1-\tilde{\dot{T}}^{2}}}.
\end{equation}
Therefore, the solution space of the rolling tachyons is unaffected by
introduction of Born-Infeld type action and is exactly the same as that
without electromagnetic fields except for rescaling of the time variable
(\ref{tim}). In addition to two trivial vacuum solutions,
$T=\pm \infty$ for $\tilde{{\cal H}}=0$ and $T=0$ for $\tilde{{\cal H}}=
\tilde{{\cal T}}_{p}$, there are three kinds of rolling tachyon solutions
as given in Eq.~(\ref{rss2})
\begin{equation}\label{rss3}
\left\{
\begin{array}{lccl}
\sinh\left(T_{{\rm RT}C}(t)/R\right)&=& \sqrt{u^{2}-1}
\cosh\left(t/\zeta\right) &
\mbox{for}~\tilde{\rho}<{\cal T}_{p} \\
\sinh(T_{\frac{1}{2}{\rm S}}(t)/R)
&=& \exp \left(t/\zeta\right)
&\mbox{for}~\tilde{\rho}={\cal T}_{p}\\
\sinh\left(T_{{\rm RT}S}(t)/R\right)
&=& \sqrt{1-u^2} \sinh\left(t/\zeta\right)
&\mbox{for}~\tilde{\rho}>{\cal T}_{p}
\end{array}
\right. ,
\end{equation}
where $u=\tilde{{\cal T}}_{p}/\tilde{\rho}\sqrt{\beta_{p}}$ and
$\zeta=\sqrt{\alpha_{p0}/\beta_{p}}R$.

The expression of energy-momentum tensor is given by symmetric part of
the cofactor $C^{\mu\nu}_{{\rm S}}$ of $(X)_{\mu\nu}$, namely,
\begin{equation}
T^{\mu\nu}=\frac{{\cal T}_{p}V}{\sqrt{-\det (X_{\mu\nu})}}C^{\mu\nu}_{{\rm S}}.
\end{equation}
For general electromagnetic fields, both momentum density $T^{0i}$
and off-diagonal stress components $T^{ij}~(i\ne j)$ do not vanish.
So we should perform Lorentz boost
transformation and spatial rotation as have been done in the case of
D3-brane~\cite{Kim:2003he} and find the frame with vanishing
$C^{ij}_{{\rm S}}~(i\ne j)$
for sufficiently large $T$. For pure electric case with $F_{ij}=0$, we
do not need such transformations since $\beta_{p}=1-{\bf E}^{2}$,
$\alpha_{p0}=1$, $C^{0i}_{{\rm S}}=0$and
$C^{ij}_{{\rm S}}=0~(i\ne j)$~\cite{Mukhopadhyay:2002en}.
If we choose the direction of electric field as $x$-axis as
${\bf E}=E\hat{{\bf x}}$, pressure
components have
\begin{eqnarray}
T_{11}&=&-\tilde{\rho}(1-\dot{T}^{2}),\\
T_{22}=\cdots =T_{pp}&=&-\tilde{\rho}(1-E^{2}-\dot{T}^{2}).
\end{eqnarray}
The equation of motion forces
$\dot{T}\stackrel{t\rightarrow \infty}{
\longrightarrow} \pm\sqrt{1-E^{2}}$ so that all the pressure components vanish
for large $T$ except the component parallel to the fluid of fundamental
strings, i.e., $T_{11}\ne 0$. Note that the time scale,
$\zeta =R/\sqrt{1-{\bf E}^{2}}$, is enlarged as the electric field
approaches critical value, $|{\bf E}|\rightarrow 1^{-}$.
Cosmological implication of this time scale change may appear as prolongation
of inflation period, which will be discussed in the next section.

\section{Cosmological Implication of U(1) Gauge Fields on Unstable D-brane}
We consider the Born-Infeld type effective action describing an unstable
D$p$-brane system of tachyon and abelian gauge field coupled to
(p+1)-dimensional gravity
\begin{eqnarray} \label{action}
S &=& \frac{1}{2\kappa^2}\int d^{p+1}x\,\sqrt{-g}\,R
\nonumber \\
 &&   -{\cal T}_p\int d^{p+1}x\,V
    \sqrt{-\det(g_{\mu\nu}+\partial_\mu T\partial_\nu T+F_{\mu\nu})}.
 \nonumber \\
\end{eqnarray}

We introduce the notations
$X_{\mu\nu}=g_{\mu\nu}+\partial_\mu T\partial_\nu T+F_{\mu\nu}$
and $X=\det(X_{\mu\nu})$.
Then the equations of motion derived from the action (\ref{action}) for
the metric, the tachyon, and the gauge field are
\begin{equation}\label{eq-metric}
G_{\mu\nu} = -\kappa^2{\cal T}_p\frac{V}{\sqrt{-X}}
    \frac{g_{\mu\lambda}g_{\nu\rho}C_{\rm S}^{\lambda\rho}}{\sqrt{-g}},
\end{equation}
\begin{equation}\label{eq-tachyon}
\partial_\mu\left(\frac{V}{\sqrt{-X}}C_{\rm S}^{\mu\nu}\partial_\nu T\right)
    +\sqrt{-X}\;\frac{dV}{dT}=0,
\end{equation}
\begin{equation}\label{eq-gauge}
\partial_\mu\left(\frac{V}{\sqrt{-X}}C_{\rm A}^{\mu\nu}\right)=0,
\end{equation}
$C_{\rm S}^{\mu\nu}$ and $C_{\rm A}^{\mu\nu}$ are symmetric
and antisymmetric parts of the cofactor of $X_{\mu\nu}$, respectively.
Note that the derivatives in Eqs.~(\ref{eq-tachyon}) and (\ref{eq-gauge})
are ordinary derivatives, not covariant derivatives.
The right-hand side of Eq.~(\ref{eq-metric}) identifies
the energy-momentum tensor to be
$T^{\mu\nu}= {\cal T}_3VC_{\rm S}^{\mu\nu}/\sqrt{gX}$.

Let us consider the time-dependent spatially homogeneous configuration
of tachyon $T(t)$ and field strength tensor $F_{\mu\nu}(t)$.
The configuration is not in general isotropic due to the
non-vanishing gauge field.
Due to Bianchi identity $\partial_{(\mu}F_{\nu\lambda)}=0$,
specifically $\partial_0F_{ij}+\partial_iF_{j0}+\partial_jF_{0i}=0$,
spatially homogeneous $F_{i0}$ imply that
$F_{ij}$ are constant in time.
In general $F_{i0}$ can have time-dependence in
time-dependent background geometry.
First, we consider $F_{ij}=0$ and the electric field directs
to $x_p$ direction, $F_{i0}=E\delta_{ip}$.
In accordance with our choice of field configuration,
we take the following metric ansatz
\begin{equation}
ds^2 = -dt^2+a(t)^2(dx_1^2+\cdots+dx_{p-1}^2)+b(t)^2dx_p^2.
\end{equation}
Then $X_{\mu\nu}$ is
\begin{equation}
X_{\mu\nu} = \left(\begin{array}{ccccc}
-1+\dot T^2 & 0 & \cdots & 0 & -E \\
0 & a^2 & \cdots & 0 & 0 \\
\vdots & \vdots & \ddots & \vdots & \vdots \\
0 & 0 & \cdots & a^2 & 0 \\
E & 0 & \cdots & 0 & b^2
\end{array}\right),
\end{equation}
and its cofactor $C^{\mu\nu}$ is given by
\begin{eqnarray}
C^{\mu\nu} &=& \left(\begin{array}{ccccc}
a^{2(p-1)}b^2 & 0 & \cdots\\
0 & -a^{2(p-2)}b^2(1-\dot T^2-E_*^2) & \cdots \\
\vdots & \vdots & \cdots \\
0 & 0 & \cdots \\
-a^{2(p-1)}E & 0& \cdots
\end{array}\right.
\nonumber \\
&& \left.\begin{array}{ccccc}
0 & a^{2(p-1)}E \\
0 & 0 \\
\vdots & \vdots \\
\hspace{-6mm}-a^{2(p-2)}b^2(1-\dot T^2-E_*^2) & 0 \\
0 & -a^{2(p-1)}(1-\dot T^2)
\end{array}\right), \nonumber \\
\end{eqnarray}
where we introduced a new variable $E_*\equiv E/b$.
For the above homogeneous configuration,
Einstein equation (\ref{eq-metric}) becomes
\begin{eqnarray}
\label{g00}
&&\frac{(p-1)(p-2)}{2}\frac{\dot a^2}{a^2}
    +(p-1)\frac{\dot a}{a}\frac{\dot b}{b}  =
    \kappa^2{\cal T}_p\,\rho, \\
\label{gii}
&&\frac{(p-2)(p-3)}{2}\frac{\dot a^2}{a^2}
    +(p-2)\frac{\dot a}{a}\frac{\dot b}{b}
    +(p-2)\frac{\ddot a}{a}+\frac{\ddot b}{b}
\nonumber \\
&& ~\hskip 3.1cm = \kappa^2{\cal T}_p\,\rho\left(1-\dot T^2-E_*^2\right), \\
\label{gpp}
&&\frac{(p-1)(p-2)}{2}\frac{\dot a^2}{a^2}+(p-1)\frac{\ddot a}{a}  =
    \kappa^2{\cal T}_p\,\rho\left(1-\dot T^2\right),
    \nonumber \\
\end{eqnarray}
where we defined the energy density function
\begin{equation}
\rho = \frac{V(T)}{\sqrt{1-\dot T^2-E_*^2}}.
\end{equation}
{}From Eqs.~(\ref{g00})--(\ref{gpp}) (or directly from ${T^{\mu\nu}}_{;\mu}=0$),
we can derive an energy-momentum conservation equation
\begin{equation} \label{emcon}
\dot\rho+\left[\left((p-1)\frac{\dot a}{a}+\frac{\dot b}{b}\right)\dot T^2
    +2\frac{\dot a}{a}E_*^2\right]\rho=0.
\end{equation}
The tachyon equation (\ref{eq-tachyon})
and the gauge field equation (\ref{eq-gauge}) are
\begin{equation}
\frac{d}{dt}(\rho\dot T)
+\left[(p-1)\frac{\dot a}{a}+\frac{\dot b}{b}\right]\rho\dot T
+\sqrt{1-\dot T^2-E_*^2}\;\frac{dV}{dT}=0,
\end{equation}
\begin{equation}
\frac{d}{dt}(\rho E)
+\left[(p-1)\frac{\dot a}{a}-\frac{\dot b}{b}\right]\rho E=0.
\end{equation}
Using Eq.~(\ref{emcon}), we rewrite them as
\begin{eqnarray}\label{tachyon}
\ddot T + \left[(p-1)\frac{\dot a}{a}\left(1-\dot T^2-E_*^2\right)
    +\frac{\dot b}{b}\left(1-\dot T^2\right)\right]\dot T
    \nonumber \\
  +\left(1-\dot T^2-E_*^2\right)\frac{1}{V}\frac{dV}{dT}
    =0,
\end{eqnarray}
\begin{equation}\label{efield}
\dot E + \left[(p-1)\frac{\dot a}{a}\left(1-\dot T^2-E_*^2\right)
    -\frac{\dot b}{b}\left(1+\dot T^2\right)\right]E=0.
\end{equation}

The distinction between the scale factors $a$ and $b$ comes from
the existence of uniform electric field along $x_p$-direction.
We can make it manifest by deriving the equation for $b_*\equiv b/a$
from Eqs.~(\ref{g00})--(\ref{gpp})
\begin{equation}
\frac{\ddot b_*}{b_*}+p\frac{\dot a}{a}\frac{\dot b_*}{b_*}=-\rho E_*^2.
\end{equation}

We try to solve the above equations
with the tachyon potential (\ref{V3})
for appropriate initial conditions
and examine the evolution of the scale factors $a$ and $b$,
the tachyon $T$ and the electric field $E$.

For numerical analysis, it is more convenient to use the variables
\begin{eqnarray}
&& a_L=\log a, \quad
b_{*L}=\log\left(\frac{b}{a}\right), \quad
\nonumber \\
&&T_*=\frac{T}{R}, \quad
E_*=\frac{E}{b}, \quad
t_*=\frac{t}{R},
\end{eqnarray}
instead of $a$, $b$, $T$, $E$ and $t$.
Using the relations $\dot a/a=\dot a_L$, $\ddot a/a=\ddot a_L+\dot a_L^2$,
 and $\dot b/b=\dot a_L+\dot b_{*L}$,
we rewrite four equations we used for numerical study
\begin{equation}
(p-1)\ddot a_L+\frac{1}{2}p(p-1)\dot a_L^2=\kappa^2{\cal T}_p R^2\,\rho(1-\dot T_*^2),
\end{equation}
\begin{equation}
\ddot b_{*L}+\dot b_{*L}+p\dot a_L\dot b_{*L}=-\kappa^2{\cal T}_pR^2\,\rho E_*^2,
\end{equation}
\begin{eqnarray}
\ddot T_* + \left[(p-1)\dot a_L\left(1-\dot T_*^2-E_*^2\right)
    +(\dot a_L+\dot b_{*L})\right.
    \nonumber \\
    \left.\left(1-\dot T_*^2\right)\right]\dot T_*
    +\left(1-\dot T_*^2-E_*^2\right)\frac{1}{V(T_*)}
    \frac{dV(T_*)}{dT_*} =0,
\end{eqnarray}
\begin{equation}
\dot E_* +\left[(p-1)\dot a_L\left(1-\dot T_*^2-E_*^2\right)
    -(\dot a_L+\dot b_{*L})\dot T_*^2\right]E_*=0.
\end{equation}

The evolution time scale is determined by two parameters
$R$ and $\kappa^2{\cal T}_pR^2$ and initial conditions.
In our numerical study, $R$ is the unit of time
and we take $\kappa^2{\cal T}_pR^2=1$ for convenience.

Let us turn to the discussion about initial conditions.
The initial state of the system are specified by three parameters
$T(0)\equiv T_i$, $\dot T(0)\equiv \dot T_i$, and $E(0)\equiv E_i$,
which fix the initial energy density to be
$\rho(0)=V(T_i)/\sqrt{1-\dot T_i^2-E_i^2}\equiv\rho_i$.
We can choose $a(0)=b(0)=1$ by coordinate rescaling.
The time derivatives $\dot a(0)$ and $\dot b(0)$ are constrained by
the Einstein equation.
Assuming that $\dot a(0)=\dot b(0)$ for simplicity,
we have $\dot a(0)=\sqrt{2\rho_i/p(p-1)}$ from Eq.~(\ref{g00}).
In terms of $a_L$, $b_{*L}$, $T_*$, and $E_*$,
above initial conditions correspond to
\begin{eqnarray}
&&a_L(0)=0,\quad \dot a_L(0)=\sqrt{2\rho_i/p(p-1)},\quad
\nonumber \\
&&b_{*L}(0)=\dot b_{*L}(0)=0,
 \\
&&T(0)=T_i,\quad \dot T(0)=\dot T_i,\quad E_*(0)=E_i.
\nonumber
\end{eqnarray}

\begin{figure}
\label{fig1}
\centerline{\includegraphics[width=75mm]{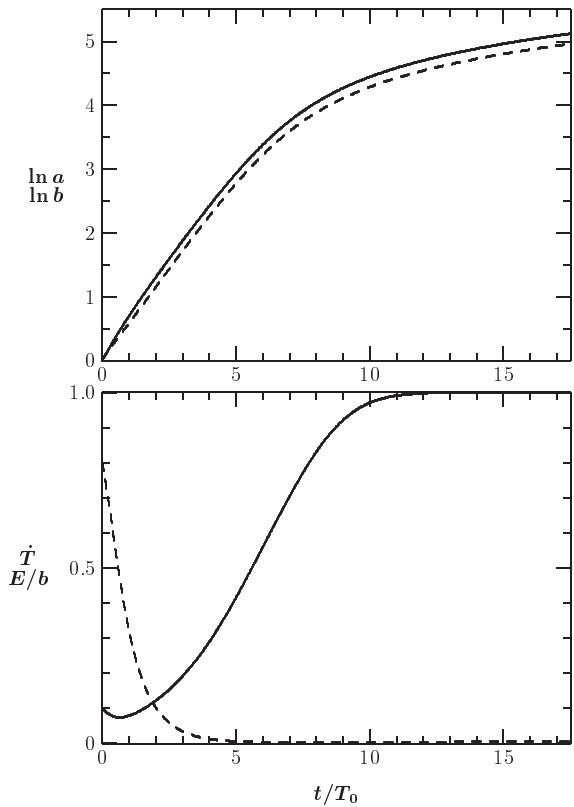}}
\caption{A numerical solution with initial conditions
$T_i=\dot T_i=0.1$, $E_i=0.8$.
In upper figure, solid and dashed lines are e-foldings
of scale factors $a$ and $b$, respectively.
In lower figure, the solid line is $\dot T$ and the dashed line is $E/b$.}
\end{figure}

We showed the numerical solutions for $p=3$ case in FIG.~1 and FIG.~2.
The general behavior is that as the universe expands
the electric field vanishes quickly
and tachyon rolling dominates in the end.
This occurs in a few e-foldings of scale factor
for moderate initial conditions, as shown in FIG.~1.
We can also think of rather extreme initial conditions.
In FIG.~2, we showed the solution for tiny value of
$T_i$ and $\dot T_i$ and nearly critical value of $E_i$.
Near the top of the potential, the existence of critical
electric field gives a rapid expansion due to the huge energy density it has.
However the electric field itself dies exponentially and
the final e-folding of scale factor seems not very significant,
unless serious fine tuning is involved.
For example, we obtained about 5 additional e-foldings in FIG.~1
from a fine tuning of $\sqrt{1-\dot T_i^2-E_i^2}=10^{-5}$,
compared to the case $E_i=0$.
The existence of critical electric field is not more helpful
than the fine tuning of initial values of tachyon field
as far as inflation is concerned.

\begin{figure}
\label{fig2}
\vspace{3mm}
\centerline{\includegraphics[width=75mm]{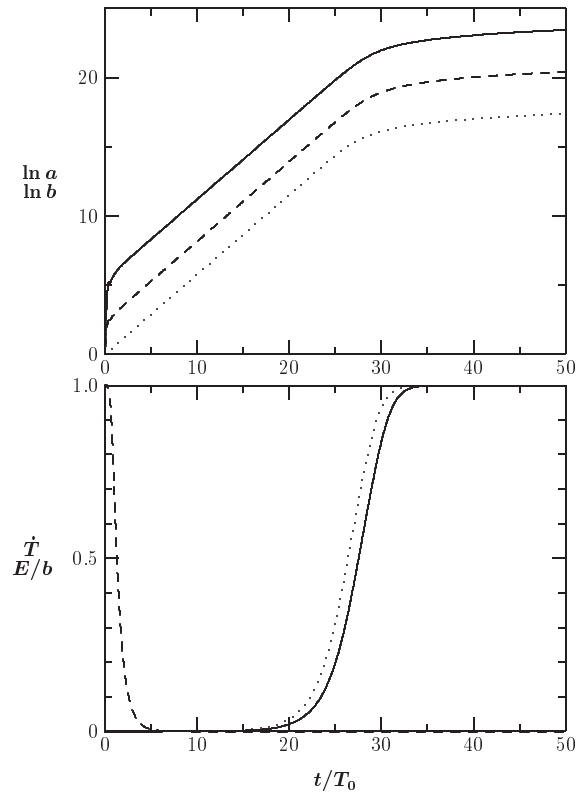}}
\caption{A solution with fine tuned initial condition
$T_i=\dot T_i=\sqrt{1-\dot T_i^2-E_i^2}=10^{-5}$.
Solid and dashed lines are same as FIG.~1.
Dotted lines are e-folding of scale factor $a$ and $\dot T$,
in the case of $E_i=0$ with same $T_i$ and $\dot T_i$.}
\end{figure}

\section{Concluding Remarks}
In this paper we have discussed a few topics of tachyon driven
cosmology based on cosmological time-dependent solutions so-called
rolling tachyons describing the homogeneous decay of unstable space-filling
D-brane. In relation with the issue of inflation, it seems unlikely to
provide sufficient e-folding by any runaway tachyon potential.
Addition of linear dilaton or Born-Infeld electromagnetic
fields is not enough to prolong sufficiently the period of inflation without
a fine-tuning.

Though we have dealt with the cosmological solutions without spatial
dependence, such solutions with both time and spatial dependence should
be studied in relation with cosmological density fluctuation.
As a viable source of density perturbation, various stable D-branes of
codimension-one or -two or composites of D-brane-fundamental string
have been proposed as static
configurations~\cite{Sen:2003tm,Lambert:2003zr,Kim:2003in}
because of the
absence of perturbative open string degrees after the decay of unstable
D-brane. Until now, on the cosmological density perturbation due to
tachyon, there exist both bad news like caustics~\cite{Felder:2002sv}
and recent good news
based on numerical analysis~\cite{Felder:2004xu}.

\acknowledgments
This work was supported by Korea Research Foundation Grants
(No. R01-2003-000-10229-0(2003) for C.K.) and is
the result of research activities (Astrophysical Research
Center for the Structure and Evolution of the Cosmos (ARCSEC))
supported by Korea Science $\&$ Engineering Foundation(Y.K., O.K., and C.L.).

\end{document}